\documentclass[aps,amsmath,amssymb,twocolumn,prl]{revtex4-2}

\usepackage{graphicx}% Include figure files
\usepackage{dcolumn}% Align table columns on decimal point
\usepackage{bm}% bold math
\usepackage{color}
\usepackage[normalem]{ulem}
\usepackage{amsmath}
\usepackage{braket}
\usepackage{hyperref}% add hypertext capabilities

\usepackage{enumitem}

\begin{document}
	\pagenumbering{arabic}
	
\title{Reflectionless modes as a source of Weyl nodes in multiterminal Josephson junctions}

%\author{David Christian Ohnmacht$^{1,\ast,\dagger}$, Valentin Wilhelm$^{1,\ast}$, Hannes Weisbrich$^{1}$, Wolfgang Belzig$^{1}$}

%\affiliation{$^{1}$Fachbereich Physik, Universit{\"a}t Konstanz, D-78457 Konstanz, Germany}

\author{David Christian Ohnmacht}
\email{david.ohnmacht@uni-konstanz.de}
\author{Valentin Wilhelm}
\author{Wolfgang Belzig}

\affiliation{Fachbereich Physik, Universit{\"a}t Konstanz, D-78457 Konstanz, Germany}
	
\begin{abstract}
Multiterminal Josephson junctions are a promising platform to study non-trivial topology in engineered quantum systems. Yet, experimentally meaningful insight into what exactly makes these systems topologically non-trivial remains elusive. In this work, we show that zero energy reflectionless scattering modes (RSMs) of the normal scattering matrix result in topological phase boundaries. By analyzing two different setups, we explain the origin of each topological phase boundary and provide generalizable insight into these systems. The considerations here can be of help for experimentalists as it connects the properties of the normal scattering region to the Andreev bound state spectrum of multiterminal superconducting junctions. In addition, our work provides a way to give meaning to the edge modes of these engineered topological systems with synthetic dimensions by providing a correspondence between unity transmission modes and boundaries between topological trivial and non-trivial phases.
\end{abstract}

\date{\today}

\maketitle
% \textbf{Topology, a branch of mathematics concerned with the properties of geometric objects that remain invariant under continuous deformations, has emerged as a powerful framework for understanding and predicting quantum phenomena in condensed matter systems [X].}

\emph{Introduction}.-- Topology manifests itself as one of the most fundamental aspects of condensed matter systems. Such systems include topological insulators \cite{Kane2005, Kane2005a, Bernevig2006, Fu2007, Konig2007}, topological superconductors \cite{Kitaev2001, Fu2008, Sato2009, Alicea2012} and quantum Hall systems \cite{klitzing1980new,TKKN,hatsugai1993chern}. Another promising direction lies in the ability to engineer the topological properties via control parameters to study topological phase transitions. Examples include topological photonics \cite{lu2014topological,wang2009observation,khanikaev2013photonic}, topological-driven Floquet systems \cite{rudner2020band,kitagawa2010topological}, topological electrical \cite{imhof2018topolectrical}, and superconducting circuits \cite{fatemi2021weyl,peyruchat2021transconductance,herrig2022cooper,Peyruchat2024a}. Furthermore, while the existence of a bulk-boundary correspondence in such topological synthetic systems is often not evident; it is still possible to infer effective boundary effects from bulk topology with time-boundary effects \cite{Xu2025} or engineered boundaries \cite{Dutt2022}. Multiterminal Josephson junctions (MTJJs) have proven to be another promising route towards engineered topology where the topology is encoded in the superconducting phase degree of freedom  \cite{riwar2016multi,eriksson2017topological,xie2017topological,xie2018weyl,deb2018josephson,xie2019topological,gavensky2019topological,houzet2019majorana,klees2020microwave,klees2021ground,Weisbrichtensor,PhysRevB.107.165301,Peralta_Gavensky_2023,riwar2019fractional,PhysRevB.107.035408,weisbrich2021second,xie2022non,Yokoyama2017, Yokoyama2015a, Meyer2017b,Repin2022,Barakov2023a}. Additionally, MTJJs also host novel (topological) phases in the context of non-hermiticity \cite{cayao2023bulk, PhysRevLett.133.086301, cayao2023nonhermitian, PhysRevB.109.214514,10.1063/5.0215522,Pino2024, cayao2024nonhermitianminimalkitaevchains,cayao2024nonhermitianmultiterminalphasebiasedjosephson, Ohnmacht2024a}.  Whereas observable consequences of the non-trivial topology are predicted to exist \cite{riwar2016multi,klees2020microwave}, whether these signatures can be observed in real systems is not clear \cite{Repin2019a}. Experimentally, recent advances in sample fabrication lead to the spectroscopic measurement of Andreev bound states (ABSs) with phase control via flux lines \cite{Coraiola2023b} which also represents an experimental realization of an Andreev molecule \cite{Pillet2019a,Kornich2019a}. In particular, phase manipulation of a four-terminal MTJJ was established experimentally in Ref.~\cite{Antonelli2025}, which satisfies the minimal requirement for the existence of Weyl nodes in MTJJs with time-reversal symmetry \cite{riwar2016multi}. Whereas the theory provided in Ref.~\cite{Antonelli2025} predicts non-trivial topology in accordance with spectroscopic data, additional experimental evidence is needed to proof the existence of non-trivial topology. 

Although MTJJs and their topology have been studied in detail for nearly a decade, making a connection from non-trivial topology to experimentally relevant parameters is difficult. Promising samples must host ABSs very close to zero energy, yet in a multiterminal setup, such a zero energy ABS is not necessarily related to a unity transmission mode as it is not possible to label ABSs by transmission eigenvalues of the full scattering matrix \cite{VanHeck2014}. Furthermore, a bulk-boundary correspondence in these systems still remains elusive, as there is no clear analog of a boundary in topological systems with synthetic dimensions.

In this manuscript, we provide a concrete explanation for the appearance of non-trivial topology and Weyl nodes in MTJJs by demonstrating that reflectionless scattering modes (RSMs) \cite{PhysRevA.102.063511} of the normal scattering matrix result in topological phase boundaries. We illustrate our findings by analyzing a four-terminal hybrid system used in Ref.~\cite{10.21468/SciPostPhys.15.5.214} where we discuss the conditions for topological phase boundaries in detail. We show that each phase boundary is related to a zero-energy RSM (zero-RSM) of the normal scattering matrix. In the second part, we analyze the four-terminal system used in Ref.~\cite{Antonelli2025} and its corresponding topological phase diagram in detail. We find that most topological phase boundaries correspond to zero-RSMs of the normal scattering matrix. Additionally, we find abnormal topological phase boundaries, which do not arise from RSMs, yet are related to effective RSMs. Additionally, we find that the source of topology is inherent to the non-trivial topology of RSMs \cite{PhysRevA.102.063511}. The findings here help to provide explanations for the occurrence of zero-energy solutions and topology in experimental MTJJ setups. Furthermore, the topological phase boundaries defined in this work provide a way to define a boundary of the synthetic bulk, thus defining an effective bulk-boundary correspondence with unity transmission modes.

%Going beyond this approximation, we discuss the characteristics of the density of states for such a junction, which reflects the real and imaginary components of the energy levels of the effective low-energy Hamiltonian for small couplings.
%Lastly, we investigate the conductivity of the system when a voltage is applied to the normal terminal, thereby providing a tool to measure the spectrum of these non-hermitian bound states.

%%%%%%%%%%%%%%%%
\emph{Reflectionless scattering modes (RSMs).--}
We consider a normal scattering region parametrized by a $n\times n$ normal scattering matrix $S_{\rm N}$, coupled to $n$ superconducting (SC) leads with phases $\bm{\phi} = {\rm diag}(\phi_1,...,\phi_n)$. Let's assume that $S_{\rm N}$ hosts a zero energy reflectionless scattering mode (zero-RSM). Then, $S_{\rm N} = \begin{pmatrix}r & t'\\t & r'\end{pmatrix}$ with the $m\times m$ ($m<n$) reflection matrix $r$ such that $\det(r^0) = 0$ with $r^0 = r(E=0)$ \cite{PhysRevA.102.063511}. We choose the SC phases such that the $m$ phases corresponding to the relevant states of the reflection matrix $r$ are zero, whereas the other $n-m$ phases are $\pi$. Then, it holds that $e^{\imath \bm{\phi}} = {\rm diag}( I_{m} ,-I_{n-m})$ with the $k\times k$ identity matrix $I_k$. Using Beenakker's determinant formulas \cite{PhysRevLett.67.3836} at zero energy, $\det(\{S_{\rm N}^0, e^{\imath \bm{\phi}} \}) = 0$ with $S_{\rm N}^0 = S_{\rm N}(E = 0)$, assuming time-reversal symmetry $S_{\rm N} = S_{\rm N}^T$,
we find the simple expression $\det {\rm diag}(r^0, -{r'}^0) =0$ 
which is fulfilled by assuming a zero-RSM. In addition, a RSM causes a transmission eigenvalue, i.e. an element of the spectrum $\sigma(tt^\dagger)$, to be unity. Thus, in the presence of a zero-RSM, the MTJJ can be treated as an effective two terminal Josephson junction where a zero mode is reached by having a phase-shift of $\pi$ between the leads and unit transmission like is the case for a general two-terminal point contact \cite{PhysRevLett.67.3836}.

\begin{figure*}
    \centering
    \includegraphics[width=1\linewidth]{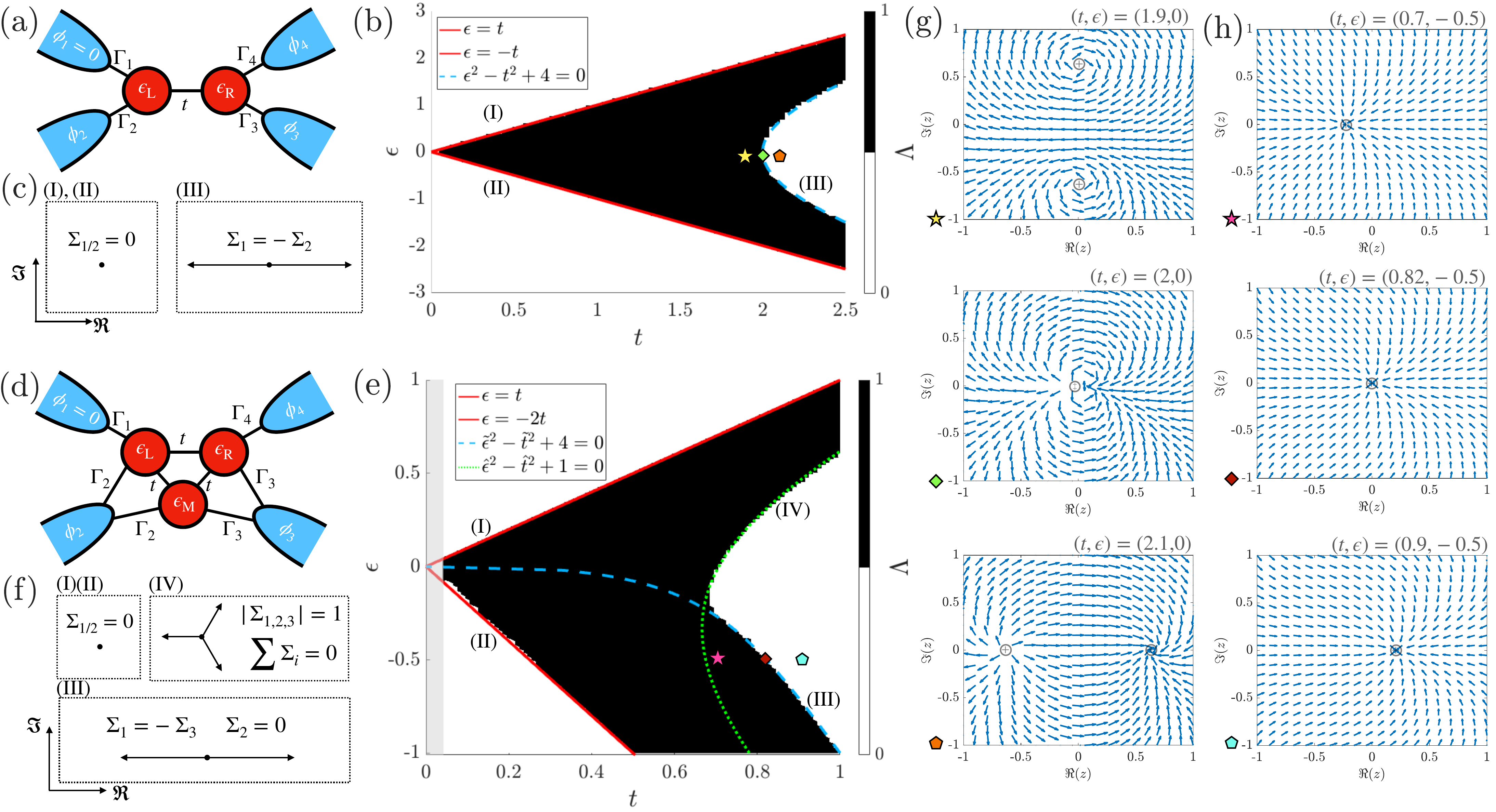}
    \caption{(a) Four-terminal MTJJ with two dots with energies $\epsilon_{\rm L/R}$ coupled to four SC terminals with phases $\phi_{1,2,3,4}$ via couplings $\Gamma_{1,2,3,4}$. (b) Topological index $\Lambda(\epsilon,t)$ indicating which parts in parameter space may host non-trivial topology. Red solid lines (I) and (II) correspond to zero-RSMs of reflection matrix $r_{13}$ with $\epsilon \pm t= 0$ ($E_\pm = 0$). The dashed line (III) corresponds to another zero-RSM of the reflection matrix $r_{12}$ with $\epsilon^2-t^2+4 = 0$ ($E_+E_-+4 = 0$). (c) Self-energies $\Sigma_{1/2}$ in the complex plane for the topological phase boundaries (I) and (II) and for the non-trivial case (III). (d) Four-terminal MTJJ with three dots with energies $\epsilon_{\rm L/M/R}$ coupled to four SC terminals with phases $\phi_{1,2,3,4}$. (e) Topological index $\Lambda(\epsilon,t)$ as in panel (b). Red solid lines (I) and (II) correspond to zero-RSMs of the reflection matrix $r_{145}$ with $\epsilon - t= 0 $ ($E_- = 0$) and $\epsilon +2t= 0$ ($E_+ = 0$). The dashed line (III) corresponds to the zero-RSM of $r_{123}$ with $\tilde{\epsilon}^2-\tilde{t}^2+4 = 0$. The dotted line (IV) corresponds to the topological phase boundary following $\hat{\epsilon}^2-\hat{t}^2+1 = 0$. (f) Self-energies $\Sigma_{1/2/3}$ in the complex plane for the zero-RSMs (I) and (II), for the non-trivial zero-RSM (III) and for the boundary not corresponding to a zero-RSM (IV). (g) The vectorfield $\bm{V}_{12}$ in Eq.~\ref{vec} of the reflection matrix $r_{12}(z)$ with complex energy $z$ for the parameter configurations seen in the insets and in panel (b). (h) The vectorfield $\bm{V}_{123}$ of the reflection matrix $r_{123}(z)$ for the parameter configurations seen inthe insets in panel (e).}
    \label{fig1}
\end{figure*}

\emph{2-dot junction}.-- We consider a double quantum dot junction that is tunnel coupled to four superconducting leads \cite{10.21468/SciPostPhys.15.5.214}, as depicted in Fig.~\ref{fig1}(a). 
Such models are used in the context of quantum dot physics \cite{martin2011josephson,recher2001andreev,governale2008real,pala2007nonequilibrium,eldridge2010superconducting,futterer2009nonlocal,herrmann2010carbon,tanaka2010correlated,tanaka2008andreev,jonckheere2009nonequilibrium,de2010hybrid}, or as a tool to parametrize a scattering region with a few high-energy ABS levels \cite{Meyer2017b,klees2020microwave,klees2021ground,Coraiola2023b,Ohnmacht2024,Ohnmacht2024a,Antonelli2025}. We assume symmetry in the system, meaning that the on-site energies of the quantum dots and couplings are equal, $\epsilon \equiv \epsilon_{\rm L/R}$ and $\Gamma \equiv \Gamma_i$ for $i = 1,2,3,4$. We choose the phases such that $\phi_1 = 0$, $\bm{\phi} = {\rm diag}(0,\phi_2,\phi_3,\phi_4)$ and $\Gamma = 1$. The low-energy physics of the system is described by an effective (topological) Hamiltonian \cite{Wang2012, Wang2013, Gavensky2023}
\begin{align}\label{topHam}
	H_{\rm eff} = \begin{pmatrix}
		H_{\rm N} & \Sigma\\  \\ \Sigma^* & -H_{\rm N}
	\end{pmatrix},
\end{align}
with the normal Hamiltonian $H_{\rm N} = \epsilon \sigma_0 + t \sigma_x$ with the Pauli matrices $\sigma_i$ and the self-energy 
\begin{equation}\label{self1}
\Sigma = \begin{pmatrix}
    \Sigma_1 & 0 \\ 0 &  \Sigma_2 
\end{pmatrix} = \begin{pmatrix}
    1+e^{\imath \phi_2} & 0 \\ 0 & e^{\imath \phi_3}+e^{\imath \phi_4}
\end{pmatrix}
\end{equation}
determined by the interconnection geometry. The Chern number $C_{i}^{jk}$ of the $i$th band can be defined
\begin{equation}\label{Chern}
    C_{i}^{jk} = \frac{1}{2\pi} \int_{0}^{2\pi}\int_{0}^{2\pi} B_i^{ jk} \, d\phi_{j} d\phi_{k},
\end{equation}
where $j\neq k$ and the total Chern number $C^{jk} = \sum_{i} C_i^{jk}$ as the sum over all occupied bands with the Berry curvature $B_i^{jk} \equiv -2{\rm Im}\braket{\partial_{\phi_{j}} v_i| \partial_{\phi_{k}} v_i}$ of the $i$th eigenvector of $H_{\rm eff}$, $\ket{v_i}$. We define the topological index $\Lambda$
\begin{equation}\label{Lambda}
    \Lambda(\epsilon,t) = \begin{cases} 0, \ C^{jk} \equiv 0  \mathrm{\ for \ all \ } j\neq k  \\ 1, \ \mathrm{otherwise}\end{cases},
\end{equation}
which is a measure for whether non-trivial topology can be achieved depending on the system parameters $\epsilon$ and $t$. In other words, if $\Lambda = 1$, there is a finite non-trivial topological region in SC phase space $\{ \bm{\phi}\}$. Furthermore, we notice that a boundary between a topologically non-trivial ($\Lambda\neq 0$) and trivial region ($\Lambda = 0$) can be viewed as a boundary of a bulk topological phase. 

In Fig.~\ref{fig1}(b), we show the topological phase diagram $\Lambda(\epsilon,t)$ of the 2-dot junction which was calculated numerically using Ref.~\cite{Fukui2005}. We observe three phase boundaries, indicated by the two straight lines (I) and (II) corresponding to the relation $\epsilon = \pm t$, and the curved line (III) with $\epsilon^2 -t^2+4= 0$. %Furthermore, for these phase boundaries, we find conditions for the eigenvalues of normal Hamiltonian $H_{\rm N}$ ($E_i$) and eigenvalues of the normal metal scattering matrix $S_{\rm N}^0$ ($d_i$). 
%We find that
%\begin{align}
%   {\rm (I)\&(II)} & \epsilon &= \pm t  &\Leftrightarrow&  E_i &=  0   
% &\Leftrightarrow&  d_i  &=  -1 \\
%    {\rm (III)} & \epsilon^2-t^2 +4 &=  0   &\Leftrightarrow&   E_iE_j +4 &=0  & \Leftrightarrow &  d_i & = -d_j.
%\end{align}
In particular, we find the following three aspects concerning the topological phase boundaries, i.e. they arise when:
\begin{itemize}[noitemsep,topsep=0pt]
    \item There is a zero-RSM in a reflection matrix $r$ of $S_{\rm N}$ such that $\det(r^0) = 0$.
    \item One eigenvalue of $H_{\rm N}$ is zero ($E_\pm = \epsilon\pm t = 0 $) or coupling terms are zero ($E_+E_-+4 = 0$).
    \item The trace of the self-energy is zero (${\rm tr}(\Sigma)= 0$).  
%    \item The convex hull spanned by the eigenvalues $d_i$ of $S_{\rm N}^0$ crosses zero.
\end{itemize}
In the following, we highlight the inter-connectivity between all of these aspects. 

First, we consider the case where the self-energy in Eq.~\ref{self1} is zero, $\Sigma = 0$. This is the case when $\bm{\phi} = (0,\pi,\phi,\phi+\pi)$ with $\phi \in [0,2\pi)$. Then, it follows that $\det(H_{\rm eff}) = \det(H_{\rm N})^2$ which is zero exactly when an eigenvalue of $H_{\rm N}$ is zero. However, how is that related to a RSM? Note, that the determinant of the reflection matrix $r$ can generally be written as a functional of $H_{\rm N}$ and the incoming $\Sigma_{\rm in}$ and outgoing $\Sigma_{\rm out}$ self-energies \cite{PhysRevA.102.063511}. In particular, for the $2 \times 2$ reflection matrix of lead 1 and 3, $(r_{13})_{ij}= (S_{\rm N})_{ij}$ for $i,j= 1,3$, it holds that
\begin{align}\label{r1}
    \det(r_{13}) & = \frac{\det(E-H_{\rm N} -\imath (\Sigma_{\rm in}-\Sigma_{\rm out}))}{\det(E-H_{\rm N}+\imath \Sigma_0)} \\
    \Rightarrow \det(r_{13}^0)& = \frac{\det(H_{\rm N})}{\det(H_{\rm N}-\imath \Sigma_0) }
\end{align}
where $\Sigma_{\rm in}= \Sigma_{\rm out} = \Sigma_0/2$ and $\Sigma_0 = \Sigma_{\bm{\phi} = 0}$. It follows that $\det(H_{\rm eff}) = 0 \Leftrightarrow \det(H_{\rm N}) = 0 \Leftrightarrow \det(r_{13}^0) = 0$, exactly when $E_\pm= \epsilon \pm t =  0$. Most interestingly, we find a relation between effective phase dependent self-energies $\Sigma$ [Eq.~\ref{self1}] and zero-RSMs of the reflection matrix $r_{13}$ which does not itself depend on the superconducting phases. To understand this connection, consider the following in the normal state: No phase shift is assigned to incoming modes ($\Sigma_{\rm in}$) whereas a phase shift of $\pi$ is assigned to outgoing modes ($-\Sigma_{\rm out}$) as they leave the scattering region. Incoming and outgoing modes can be understood as gain and loss states, which in the language of non-hermitian theory is accounted for by the sign of the self-energy. This assignment of different phases is exactly analogous to the effect of SC couplings on the system, namely when the phases are varied to a phase difference of $\pi$. Thus, in symmetric junctions, zero-RSMs effectively encode the same information as effective two-terminal Josephson junctions which are tuned to a $\pi$ phase difference. 

Now, let's consider the case where ${\rm tr}(\Sigma) =0$ with $\Sigma_1 = -\Sigma_2\neq 0$ which results in the condition $\bm{\phi}=(0,\phi,\pi,\phi+\pi)$ for $\phi\in [0,2\pi)$. It holds that $-\Sigma_2 = \Sigma_1 = \gamma e^{\imath \alpha}$ for $\gamma\geq 0$ and $\alpha\in [0,2\pi)$. In the case of $\gamma = 2$ [$\bm{\phi}=(0,0,\pi,\pi)$], it holds $\Sigma = 2\sigma_z$ and 
\begin{align}\label{magic}
\det(H_{\rm eff}) &=  \det \begin{pmatrix}
    H_N & 2\sigma_z \\ 2\sigma_z & -H_N
\end{pmatrix} \\  &= \det(H_N+2\imath\sigma_z)\det(-H_N+2\imath \sigma_z)
\end{align}
which yields the condition $\epsilon^2-t^2+4= 0$ or equivalently $E_+E_-+4 = 0$. Like before, we find the corresponding reflection matrix, which in this case is the $2\times 2$ matrix corresponding to terminal 1 and 2, $(r_{12})_{ij}= (S_{\rm N})_{ij}$ for $i,j = 1,2$, where the incoming and outgoing self-energies read $\Sigma_{\rm in} = \sigma_0+\sigma_z$ and $\Sigma_{\rm out} = \sigma_0-\sigma_z$ resulting in
\begin{equation}\label{r2}
    \det(r_{12}) = \frac{\det\left(E-H_{\rm N}-2\imath \sigma_z \right)}{\det(E-H_{\rm N}+\imath \Sigma_0)}.
\end{equation}
It is then evident that $\det(H_{\rm eff}) = 0 \Leftrightarrow \det(r_{12}^0) = 0$. As before, the reflection matrix imitates the effect of SC phases by assigning different phases to incoming and outgoing states respectively. In the previous case, incoming and outgoing states are mediated via both dots, meaning that $\Sigma_{\rm in} = \Sigma_{\rm out}$, which effectively eliminates any contribution from the self-energy in the numerator in Eq.~\ref{r1}. In this case, incoming and outgoing states see different self-energies, resulting in a non-trivial non-hermitian contribution in the numerator of Eq.~\ref{r2}. Hence, we get a more involved relation involving both energy levels granting a more complex phase boundary as seen in Fig.~\ref{fig1}(b), (III). Furthermore, zero-RSMs are tied to geometric rules of the self-energies. This enables us to assign to each phase boundary observed in Fig.~\ref{fig1}(b) a respective diagram like it is seen in Fig.~\ref{fig1}(c).

Whereas the former considerations show that zero-RSMs provide ABSs at zero energy, it is not evident why they provide topological transitions. In order to demonstrate the relation to topology, consider a so-called reflection zero (R-zero) at a complex energy $z_0 \in \mathbb{C}$, meaning that $\det(r(z_0))=0$. Such a R-zero provides a topological charge via a winding number in complex energy space \cite{PhysRevA.102.063511}. To visualize this, we define the normalized vector field
\begin{equation}\label{vec}
    \bm{V}_{\alpha} = \left(\frac{{\rm Re} \det(r_{\alpha}(z))}{|\det(r_{\alpha}(z))|},\frac{{\rm Im} \det(r_{\alpha}(z))}{|\det(r_{\alpha}(z))|}\right)
\end{equation} 
of the reflection matrix $r_{\alpha}(z)$ with complex energy $z \in \mathbb{C}$. In Fig.~\ref{fig1}(g), we portray $\bm{V}_{12}$ of the reflection matrix $r_{12}(z)$ for the parameter configurations shown in the insets and also indicated in panel (b). For the left panel ($t = 1.9$, $\epsilon = 0$), it is seen that two R-zeros lie on opposite complex conjugated energies. The "+"-symbols indicate topological charges resulting in a non-trivial winding number. In the middle panel ($t = 2$, $\epsilon = 0$), at the phase transition, two R-zeros combine to one zero-RSM at zero real energy, forming an exceptional point of the effective non-hermitian operator $z-H_{\rm N}-2\imath \sigma_z$. For the right panel ($t= 2.1$, $\epsilon = 0$) the single zero-RSM splits into two RSMs going away from zero real energy. We find that a topological phase boundary arises exactly when the topological singularity of R-zeros aligns with real zero energy.

\emph{3-dot junction}.-- In the following, we consider a three quantum dot junction depicted in Fig.~\ref{fig1}(d). Notably, this system was used in Ref.~\cite{Antonelli2025} to analyze spectroscopic data of a four-terminal MTJJ. In addition, it was found in Ref.~\cite{Antonelli2025} that the model predicts non-trivial topology in a specific parameter regime. In the following, we show that we can attribute zero-RSMs of the normal scattering matrix to certain topological phase boundaries in the system. In addition, we find abnormal topological phase boundaries which have another, yet related, origin. The system is described by an effective Hamiltonian like in Eq.~\ref{topHam} with a normal Hamiltonian and the self-energy matrix
\begin{align}
H_{\rm N } &= \begin{pmatrix}
		\varepsilon & t & t \\ t & \varepsilon & t \\ t & t & \varepsilon
	\end{pmatrix}, \\ 
    \Sigma &= {\rm diag}(1+e^{\imath \phi_2},e^{\imath \phi_2}+e^{\imath \phi_3},e^{\imath \phi_3}+e^{\imath \phi_4}). \label{selff}
\end{align}
In analogy, we can compute the topological index $\Lambda$ [Eq.~\ref{Lambda}] by computing the Chern numbers [Eq.~\ref{Chern}]. 

In Fig.~\ref{fig1}(e), we show the topological phase diagram $\Lambda(t,\epsilon)$ of the junction. Two topological phase boundaries are easily observed as they correspond to straight lines following the relation $\epsilon = t$ (I) and $\epsilon = -2t$ (II). Similar to before, this corresponds to the eigenvalues of the normal Hamiltonian to be zero, namely $E_- = \epsilon-t = 0$ and $E_+ = \epsilon+2t = 0$. Likewise, this case is associated with a vanishing self-energy $\Sigma = 0$ in Eq.~\ref{selff} for $\bm{\phi} = (0,\pi,0,\pi)$. It follows that $0 = \det(H_{\rm eff}) = -\det(H_{\rm N})^2$ which is zero for $E_\pm = 0$. We find the corresponding reflection matrix $(r_{145})_{ij} = (S_{\rm N})_{ij}$ with $i,j = 1,4,5$ where incoming and outgoing self-energies cancel, which results in $\det(r_{145}^0)\propto \det(H_{\rm N})$. It follows that $\det(r_{145}^0) = 0 \Leftrightarrow \det(H_{\rm N})= 0 \Leftrightarrow \det(H_{\rm eff})= 0$.

There are two other boundaries (III) and (IV) in Fig.~\ref{fig1}(e) which cross each other, leading to a more involved phase diagram. For the dashed line (III), it holds that $\Sigma_2 = 0$ with $-\Sigma_3 = \Sigma_1 = 2$ for $\bm{\phi} = (0,0,\pi,\pi)$. We find the corresponding reflection matrix $(r_{123})_{ij}= (S_{\rm N})_{ij}$ for $i,j = 1,2,3$ with the incoming and outgoing self-energies $\Sigma_{\rm in} = {\rm diag}(2,1,0)$ and $\Sigma_{\rm out} = {\rm diag}(0,1,2)$, which results in the same effective contribution as for the phase dependent self-energies, namely $\Sigma_{\rm in}-\Sigma_{\rm out} = \Sigma(0,0,\pi,\pi) = {\rm diag}(2,0,-2)$. With a similar calculation as in Eq.~\ref{magic}, we find that $\det(H_{\rm eff}) = 0 \Leftrightarrow \det(r_{123}^0)=0$. Additionally, as the effect of SC vanishes in the second dot, we can dress it onto the two remaining dots, resulting in an effective two-dot model \cite{SM}
\begin{equation}
    \det(H_{\rm eff}) = \det{(H_{\rm eff}^{\rm 2dot})} = \det \begin{pmatrix}
        H_{\rm N}^{\rm 2dot} & \Sigma^{2dot} \\ \Sigma^{2dot} & -H_{\rm N}^{\rm 2dot} 
    \end{pmatrix},
\end{equation}
with the effective 2-dot normal Hamiltonian $H_{\rm N}^{\rm 2dot} = \tilde{\epsilon}\sigma_0 + \tilde{t}\sigma_x$ with the renormalized parameters $\tilde{\epsilon} = \epsilon-t^2/\epsilon$ and $\tilde{t} = t-t^2/\epsilon$ with the effective 2-dot self-energy $\Sigma^{\rm 2dot} = 2\sigma_z$. In particular, upon redefining parameters, we obtain the exact same condition as before for the RSM coupling two levels, namely $\tilde{\epsilon}^2-\tilde{t}^2 +4 = 0$, relating the three-dot RSM to one of an effective two-dot system. 
%Furthermore, it holds that $\tilde{\epsilon}^2-\tilde{t}^2 = (\epsilon-t)^2(\epsilon+2t)/\epsilon$.

The dotted line (IV) does not correspond to a RSM of the normal state scattering matrix. Yet, it can be inferred by constraints on the self-energies. Namely, consider the case when $\bm{\phi} = (0,\pm 2\pi/3,\mp 2\pi/3,0)$, resulting in the self-energies $\Sigma_1 = e^{\pm \imath \pi/3}$, $\Sigma_2 = -1$ and $\Sigma_3 = e^{\mp \imath \pi/3}$ with the corresponding self-energy diagram shown in Fig.~\ref{fig1}(f). As $|\Sigma_{1,2,3}| = 1$, we can use Schur's determinant formula to obtain the following formula \cite{SM}
\begin{align}
    0 = \det(H_{\rm eff}) &= \det(\Sigma+H_{\rm N}\Sigma H_{\rm N}) \\
    &= \left(1+\hat{\epsilon}^2-\hat{t}^2\right)\left(1+(\epsilon-t)^2\right)
\end{align}
where we again obtain a similar formula to the one of a RSM of a 2-level system with the renormalized parameters $\hat{\epsilon} = \epsilon+t/2$ and $\hat{t}= 3t/2$.

In Fig.~\ref{fig1}(h) we portray the vector field $\bm{V}_{123}$ of the reflectionmatrix $r_{123}$ (see Eq.~\ref{vec}) for the different parameter configurations shown in the insets and portrayed in panel (b). It is seen that each parameter configuration hosts a RSM, and for the phase transition, the RSM becomes a zero-RSM at zero real energy.

We want to emphasize that these findings are also meaningful for non-symmetric junctions. As topology results in stability over a finite parameter range, non-trivial topology remains stable upon added asymmetry. We infer that in nearly symmetric junctions, which are close in parameter space to symmetric junctions, the topology is rooted in zero-RSMs. %However, to what degree this protection is guaranteed and whether every source of non-trivial topology can always be traced back to RSMs is not clear and requires additional effort. Namely, we find phase boundaries not corresponding to RSMs, but follow very similar functional dependencies. 

\emph{Discussion.}-- In summary, we have shown that zero-RSMs result in topological phase boundaries of MTJJs. In particular, we derive formulas for every arising phase boundary in the two systems in this work, where most are related to zero-RSMs and thereby also related to effective two-terminal Josephson junctions tuned to phase differences of $\pi$. Furthermore, we have established an effective bulk-boundary correspondence for MTJJs, where we find unity transmission modes in the normal state scattering matrix on topological phase boundaries. We want to point out the similarities between our findings for MTJJs and former works \cite{Fulga2011, Beenakker2011, Pikulin2012, Fulga2012,Beenakker2013,Beenakker2013a} that associate topological phase transitions by zero eigenvalues and resonances of the reflection matrix or zero crossings and parity transitions. Further analysis could focus on computing topological phase diagrams for every possible normal state scattering matrix, but even in the simplest case of $4\times 4$ matrices, the configuration space is very large and advanced numerical methods would be required beyond the scope of this work.

In conclusion, our works provides insight into the topological properties of MTJJs. We find that zero-RSMs of the normal state lead to topological phase boundaries. In addition, our work provides substantial insight into model systems, which were recently used to explain measurements in accordance with topology. Our results could provide a useful tool in fabricating topologically non-trivial devices, by relating topology in MTJJs to the properties of the respective normal scattering matrix.

We acknowledge support by the Deutsche Forschungsgemeinschaft (DFG; German Research Foundation) via SFB 1432 (Project No. 425217212).

\bibliography{references.bib}

\newpage\phantom{blabla}

\newpage

\section{Appendix}

\emph{Analysis of the normal-state scattering matrix}.-- In the following, we provide additional information on the normal state scattering matrix and the corresponding zero energy reflectionless scattering modes (zero-RSMs). The normal state scattering matrix of the junction can by computed via the Fisher-Lee relation \cite{Fisher1981} from the retarded central Green's function $g_{\rm C} = \lim\limits_{\eta\to 0} (E-H_{\rm N} + i W W^\dagger)^{-1}$
\begin{equation}
    S_{\rm N} = 1-2i W^\dagger g_{\rm C}W,
    \label{FisherLee}
\end{equation}
with the coupling matrix $W$. 

For the two-dot model in the main text, the coupling matrix is given as 
\begin{align}
    W = \sqrt{\Gamma } \begin{pmatrix}
        1 & 1 & 0 & 0\\
        0 & 0 &1 & 1
    \end{pmatrix}.
\end{align}
Diagonalizing the normal scattering matrix ($D_S = US_{\rm N}U^{-1}$) results in 
\begin{equation}
    D_S = {\rm diag}\left(1,1,\frac{\epsilon_+^*}{{\epsilon_+}},\frac{\epsilon_-^*}{{\epsilon_-}}\right) \equiv {\rm diag} \left(1,1,d_+,d_- \right)
\end{equation}
where $\epsilon_\pm = \epsilon \pm t-2\imath \Gamma$ are the eigenvalues of the effective non-hermitian normal Hamiltonian $H_{\rm N}^{\rm nh} = (\epsilon-2\imath \Gamma) \sigma_0+t\sigma_x$. To get more insight, we can express the normal scattering matrix as a function of its eigenvalues $d_{\pm}$, $S_{\rm N} = U^{-1}D_S(d_+,d_-)U$. Now, we split the system into two parts and treat the system as an effective two terminal junction. For that, consider the reflection matrices $r_{12}$ and $r_{13}$ as defined in the main text. For a zero-RSM, one eigenvalue of the reflection matrix must be zero. Thus, we compute the diagonalized forms of the respective reflection matrices which read
%\begin{align}
%    D_{R_{12-34}} = {\rm diag}\left(1, -\frac{4\Gamma ^2+e^2-v^2}{4\Gamma ^2+4\imath\Gamma e -e^2+v^2} \right) \\
%    D_{R_{13-24}} = {\rm diag}\left( \right)
%\end{align}
\begin{align}
    D_{r_{12}} &= {\rm diag}\left(1, \frac{1}{2}(d_++d_-) \right), \\
    D_{r_{13}} &= {\rm diag}\left(\frac{1}{2}\left(1+ d_+\right),\frac{1}{2}\left(1+ d_-\right) \right).
\end{align}
We obtain zero-RSMs when $d_+ = -d_-$ for $r_{12}$ and $d_{+,-}= -1$ for $r_{13}$. Note, that these conditions relate to the convex hull spanned by the eigenvalues of the normal scattering matrix to enclose the origin on one of its borders. Furthermore, these conditions result in conditions for the eigenvalues of the effective non-hermitian Hamiltonian, ${\rm Re}(\epsilon_+{\epsilon_-}^*) = 0$ for $r_{12}$  and ${\rm Re}(\epsilon_\pm) = 0$ for $r_{13}$ which then lead to the relations involving the eigenvalues $E_{\pm}=\epsilon\pm t$ of the normal metal Hamiltonian $H_{\rm N}$, $E_+E_-+4= 0$ ($t^2 = \epsilon^2+4$) and $E_\pm = 0$ ($\epsilon = \pm t$) respectively.
As discussed in the main text, RSMs correspond to a unity transmission eigenvalue. This arises from the fact that $r$ and $r^\dagger$ share the eigenvalue 0, and consequently does the hermitian matrix $rr^\dagger$. Lastly, the matrices $1-tt^\dagger$ and $rr^\dagger$ have the same spectrum \cite{RevModPhys.69.731}. For completeness, the diagonalized form of respective transmission matrices read
\begin{align}
    T_{12-34}  &= {\rm diag}\left( 0,\frac{1}{4}\left(2-d_+{d_-}^*-{d_+}^*d_-\right)\right), \\
    T_{13-24} &= {\rm diag}\left( \frac{1}{4}\left(2-d_+-{d_+}^*\right),\frac{1}{4}\left(2-d_--{d_-}^*\right)\right),
\end{align}
and we find eigenvalues equal to 1 exactly when the same conditions as before are fulfilled.
If a transmission eigenvalue is equal to 1, we can accordingly choose the phases in a way to get a zero energy solution exactly when the phases of incoming modes are zero and the ones of the outgoing modes are $\pi$. Hence, we find that we get zero energy ABSs when RSMs (or transparent modes) appear in the normal scattering region.

For the 3-dot junction the normal state scattering matrix can be computed using the Fisher-Lee relation with a coupling matrix 
\begin{align}
   W =  \sqrt{\Gamma}\begin{pmatrix}
        1& 1& 0  & 0 & 0  & 0\\
        0& 0& 1  & 1 & 0  & 0\\
        0& 0& 0  & 0 & 1  & 1\\
    \end{pmatrix}
\end{align}
like in Eq.~(\ref{FisherLee}). Since no crossed Andreev reflections are assumed in the model, the normal state scattering region is described by a $6\times 6$ matrix. As before, the S-matrix can be diagonalized which yields
\begin{align}
    D_S &= {\rm diag}\left(1,1,1,\frac{\epsilon_+^*}{{\epsilon_+}},\frac{\epsilon_-^*}{{\epsilon_-}},\frac{\epsilon_-^*}{{\epsilon_-}}\right) \\ &= {\rm diag}\left(1,1,1,d_+,d_-,d_-\right)
\end{align}
where the eigenvalues of the effective non-hermitian central Hamiltonian 
\begin{equation}
    H_{\rm N}^{\rm nh} = \begin{pmatrix} \epsilon-2\imath \Gamma & t & t \\ t & \epsilon-2\imath \Gamma & t \\ t & t & \epsilon -2\imath \Gamma \end{pmatrix}
\end{equation}
are $D_{H_{\rm N}^{\rm nh}} = {\rm diag}(\epsilon_+,\epsilon_-,\epsilon_-)$ with $\epsilon_+ =\epsilon +2t-2\imath \Gamma$ and $\epsilon_- = \epsilon-t-2\imath \Gamma$. 

In this system, we find three zero-RSMs. The first two appear in the reflection matrix $r_{145}$, where we find
\begin{equation}
    D_{r_{145}} = {\rm diag}\left(\frac{1}{2}\left(1+ d_+\right),\frac{1}{2}\left(1+ d_-\right),\frac{1}{2}\left(1+ d_-\right) \right),
\end{equation}
in complete analogy to before. Namely, when an edge of the convex hull of the eigenvalues of the normal state scattering matrix lies on the origin, we get zero energy solutions. Another zero-RSM is found in the reflection matrix $r_{123}$, which doesn't provide any simple geometric relation for the eigenvalues of the scattering matrix.

\emph{Details on 3-dot system}.-- As mentioned in the main text, for certain topological phase boundaries, the self-energy of a specific dot may be zero. This effectively decouples the dot from its environment at a zero crossing which enables us to dress this dot onto the remaining dots. Therefore, consider the phase boundary (III) in the 3-dot junction of the main text which has vanishing self-energy on the middle dot. For a zero solution of the effective Hamiltonian to appear in Eq.~(\ref{topHam}), we infer by Schur's formula a zero energy solution of the descending Hamiltonian 
\begin{align}
    H_{\rm eff}^{\rm 2dot} = \begin{pmatrix}
        H_{\rm L } & t\sigma_3 \\ t\sigma_3 & H_{\rm R}
    \end{pmatrix}  - \frac{t^2}{\varepsilon} \begin{pmatrix}
        1 & 1 \\ 1 & 1
    \end{pmatrix} \otimes \sigma_3.
\end{align}
This corresponds to a renormalization of the dot energy and hopping parameter by $\varepsilon \to\tilde{\varepsilon} = \varepsilon - t^2/\varepsilon$ and $t\to \tilde{t} =  t - t^2/\varepsilon$, resulting in an analogous treatment as for the 2-dot junction. Thus, from the relation $4\Gamma^2 + \tilde{\varepsilon}^2 - \tilde{t}^2 =0$, the explicit form of the phase boundary reads
\begin{align}
    \varepsilon =& \frac{-\tilde{t}^2+(2\tilde{t}-\tilde{\varepsilon})\tilde{\varepsilon}}{2\tilde{t}-\tilde{\varepsilon}},\\
    t =& \sqrt{\varepsilon(\varepsilon -\tilde{\varepsilon})},
\end{align}
resulting in the more complicated from as seen in the main text.

In the symmetric 3-dot model, for the phase boundary (IV), one has (up to reordering and a global sign) the situation 
\begin{align*}
	H_{\rm N } = \begin{pmatrix}
		\varepsilon & t & t \\ t & \varepsilon & t \\ t & t & \varepsilon
	\end{pmatrix} ,\quad \Sigma = \begin{pmatrix}
	1 & 0 & 0\\ 0 & \zeta & 0 \\ 0 &0&  \zeta^2
	\end{pmatrix},
\end{align*}
with $1, \zeta, \zeta^2$ the roots of $z^3 -1 = 0$. For a zero energy ABS to appear, we find
\begin{align}
	\det\left(\Sigma + \frac{1}{\Gamma^2}H_{\rm N}\Sigma H_{\rm N}\right) = 0,
	\label{eq:benakker_dot_hamiltonian}
\end{align} 
which reduces to
\begin{align*}
	\det \begin{pmatrix}
		a & b & b^* \\ b^* & a & b\\ b & b^* & a 
			\end{pmatrix} &= a^3  + 2\mathrm{Re}(b^3) - 3 |b|^2a =  0 \\
			&= C^2(\Gamma^2 + (\varepsilon - t)^2),
\end{align*}		
where we used the following abbreviations
\begin{align*}
	a &= \Gamma^2  + \varepsilon^2  - t^2 = C + t(t-\varepsilon),\\
	b &= \zeta^2 t(t-\varepsilon),\\
	C &= \Gamma^2 + \varepsilon^2 +\varepsilon t - 2t^2 = \Gamma^2 + \left(\varepsilon +\frac{t}{2}\right)^2 - \left(\frac{3}{2}t\right)^2.
\end{align*}
This makes it clear that the condition $C = 0 $ has to be fulfilled for a zero energy solution to appear. Interestingly, this could be inferred from a two dot model with renormalized symmetric levels $\hat{\varepsilon} = \varepsilon+ t/2$ and hopping $\hat{t} = 3t/2$ that has the same two distinct eigenvalues as the central three-dot Hamiltonian. In such a two-dot geometry, the topological phase boundary would be related to a zero-RSM of the normal scattering matrix. 

\end{document}